# Electrostatic Influence of Diamond Heat-Spreaders on GaN FET Performance

Lincoln D Hogan[2], Md Mazharul Islam[1], *Graduate Student Member, IE*EE, and Ahmedullah Aziz[1*], Senior Member, *IEEE*

[1]Dept. of Electrical Eng. and Computer Sci., University of Tennessee, Knoxville, TN, 37996, USA

[2]Dept. of Electrical and Computer Eng., North Carolina State University, Raleigh, NC, 27695, USA

*Corresponding Author. Email: aziz@utk.edu

***Abstract*— Diamond, with its exceptional thermal conductivity, is widely explored as a heat spreader to improve thermal management in GaN-based power and RF electronics. In this work, we use Sentaurus TCAD to model AlGaN/GaN HEMTs with integrated diamond heat spreaders, incorporating advanced physical models such as thermal boundary resistance (TBR) and temperature-dependent bandgap narrowing. While diamond significantly enhances heat dissipation, our simulations reveal an important but often overlooked tradeoff: electrostatic modulation at the diamond and semiconductor interface. We initially modeled a p-GaN HEMT with all-around diamond integration and show agreement to the experimentally reported electrothermal benefits of diamond heat spreader integration. However, we observe strong band bending at the diamond/GaN interface alongside electric field generation and increased valence band energy in diamond. As a result, an interfacial hole accumulation region forms at the GaN surface, altering vertical transport and hindering current injection which prevents high on-state current. These results indicate that although diamond improves thermal performance, it can also actively influence the electrostatics of the transistor. To address both the elevated TBR and the undesired interfacial band bending, we propose the incorporation of an engineered interlayer. Building on our Sentaurus TCAD results, we develop a Silvaco TCAD TLM test structure to characterize the extent of the hole accumulation region formation and give insight to future studies on mitigation through interlayer development.**

## I. Introduction

Field--effect Field-effect transistors (FETs) are widely utilized in high-mobility electronic applications, including RF systems and power switching devices [1]. One of the key challenges in such high-performance environments is self-heating, which leads to elevated channel temperatures [2], [3]. This thermal buildup can degrade device performance through increased carrier scattering, mobility degradation, and increased channel resistance [3], [4]. As a result, devices experience decreased on-state current, elevated on-resistance, and adverse impacts on breakdown voltage characteristics [3], [4].

To mitigate these effects, heat spreaders composed of materials with high thermal conductivity are often integrated into the device architecture [5], [6]. Diamond offers exceptional thermal conductivity, making it an ideal candidate for high-power FETs [6], [7]. Prior studies have demonstrated that diamond integration can lower operating temperatures, thereby enhancing both on-state current and breakdown voltage, especially under high-temperature conditions [3], [6].

However, in this work, we show that diamond heat spreaders may also introduce unintended electrostatic effects. We investigate how the presence of diamond layers on GaN influence band bending and the formation of an interfacial hole-accumulation layers [8], [9]. These effects are especially relevant in short-channel devices, where strong electric fields and carrier confinement are sensitive to interface conditions [10],[11],[12]. Our findings highlight a tradeoff between thermal management and electrostatic integrity in diamond-integrated GaN-based FETs.

## II. I-V Analysis of Diamond Integrated Sentaurus TCAD HEMT

We begin by replicating the experimentally shown thermal benefits of diamond integration through numerical simulation of a p-GaN HEMT structure incorporating diamond heat spreaders [1], [6]. A two-dimensional device model was implemented in Sentaurus TCAD to evaluate the influence of diamond on device electrothermal behavior and current–voltage characteristics. Although diamond integration has been primarily demonstrated in NFET GaN HEMTs [1], p-GaN HEMTs provide a more sensitive platform for studying the electrical consequences of diamond implementation. Unlike conventional NFETs, whose strong polarization-induced 2DEG can mask subtle electrostatic perturbations, p-GaN HEMTs are highly sensitive to interface charge and band bending, enabling clearer isolation of diamond-induced electrical effects from its thermal advantages.

Baseline simulations were performed prior to diamond integration and subsequently repeated for the diamond-integrated device. The simulated PFET was calibrated to reproduce select I–V characteristics reported in [1], providing a validated reference for comparison with the NFET structure.
This agreement establishes a consistent simulation framework for evaluating diamond integration, ensuring that observed changes in electrical behavior can be attributed to diamond-induced electrostatic and thermal effects rather than discrepancies in device calibration.

Implementation of diamond in Sentaurus required the introduction of custom material definitions, as diamond is not included as a native material within the standard library [13]. Key electrical properties were defined using values reported in literature to maintain physically consistent band alignment and interface behavior [8], [7].

Temperature-dependent bandgap narrowing was implemented using Varshni's relation, and avalanche generation was modeled using the van Overstraeten–de Man formulation to capture high-field carrier multiplication during breakdown conditions [13]. These models were coupled with Sentaurus hydrodynamic transport to account for carrier heating and nonlocal energy transport, enabling accurate simulation of device behavior under elevated thermal stress [2], [3], [14].

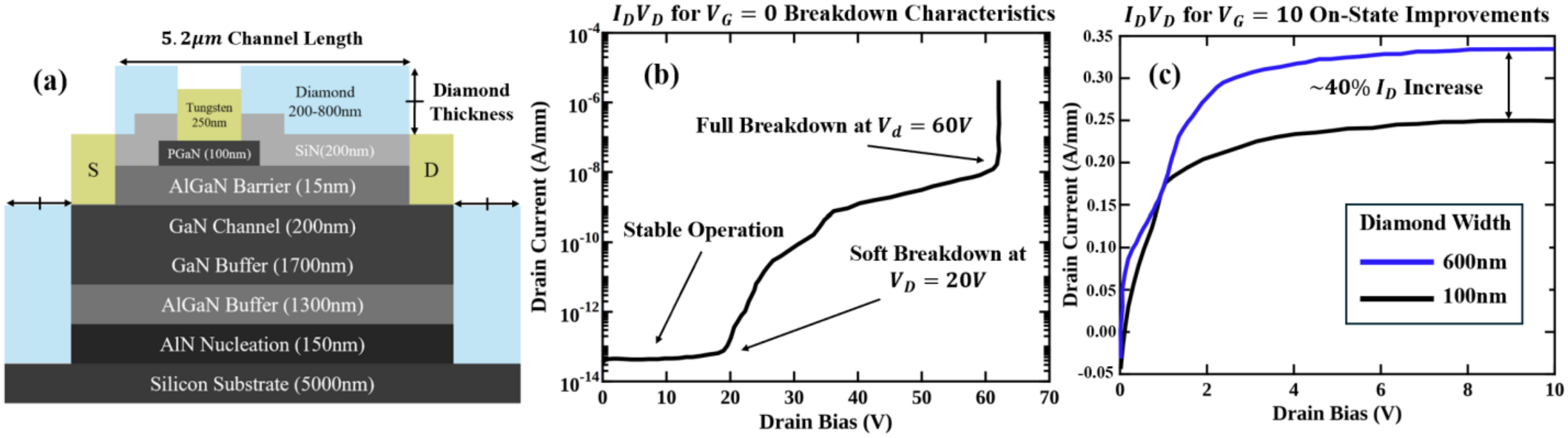

**Fig. 1: (a)** Our simulated p-GaN stack with diamond thickness and channel lengths indicated by the bidirectional arrows. **(b)** $I_D V_D$ breakdown current of 400nm diamond incorporated device during OFF operation ($V_G$=0). Soft breakdown occurred at 20V, and hard breakdown occurred at 60V, comparable to the breakdown observed in experimental literature [1]. **(c)** $I_D V_D$ at 500K with varied diamond thickness shows significant improvement in On-state current as diamond thickness increases.

Incorporation of these physical models allowed benchmarking of the diamond-integrated structure against the experimentally reported device [1], yielding comparable breakdown characteristics and validating the electrothermal simulation approach.

An effective Thermal Boundary Resistance (TBR) methodology was developed using custom material definitions. Ultrathin interlayers (1 nm) were introduced between the device and diamond regions with engineered thermal conductivity chosen to reproduce the desired interfacial thermal resistance [4], [5]. The interlayer material preserves the electrical properties of diamond while modifying only thermal transport parameters, thereby introducing a controlled thermal resistance without perturbing device electrostatics.

## III. Effects of the Increased Hole Accumulation at the Diamond/GaN Interface

While the I–V characteristics in Fig. 1(b, c) supported the reported thermal benefits of the diamond heat spreader, the observed cross-sectional carrier distributions and hole-current trends revealed behavior that could not be explained solely by thermal effects. Upon further investigation, a built-in electric field approaching 1 MV/cm was observed at the diamond/buffer interface, extending into the surrounding device layers. To evaluate the influence of this field, the device was first simulated under equilibrium conditions with no applied bias. This approach isolates the electrostatic effects introduced by diamond integration from bias-induced carrier redistribution [8]. We observe pronounced upward bending of the valence band near the diamond/buffer interface, resulting in a hole confinement region. This band curvature promotes hole accumulation while suppressing electron occupation, supporting the formation of the hole accumulation layer at the interface [9], [10].

To validate the presence and impact of the interfacial hole accumulation, we examined the hole current density. We observe a hole current density of approximately 6×10^(-2) A/cm, localized along the interface, with vectors clearly oriented downward into the substrate. This strongly indicates the presence of the hole accumulation region at the interface, which is actively contributing to leakage current in the device [9], [10], [15]. Such leakage may result in the depletion of holes from the source and drain regions, and from the buffer layer near the interface, potentially degrading device performance and short-channel control [3], [15].

To assess the impact of this leakage on device performance, we apply a drain bias and examine hole concentrations across the structure. The primary concern is that hole depletion near the source and drain contacts driven by leakage into the substrate may impede carrier injection and reduce overall current output [15]. To investigate this, we plot the hole density under a 60 V drain bias, focusing on the source region. The results in Fig. 2(c) showed that the portion of the source contact adjacent to the diamond interface exhibited a near-zero-hole concentration, with a gradient increasing toward the upper edge of the contact. This localized depletion reduces the availability of holes for injection, thereby limiting the current that can enter through the source contact and degrading the device's output characteristics.

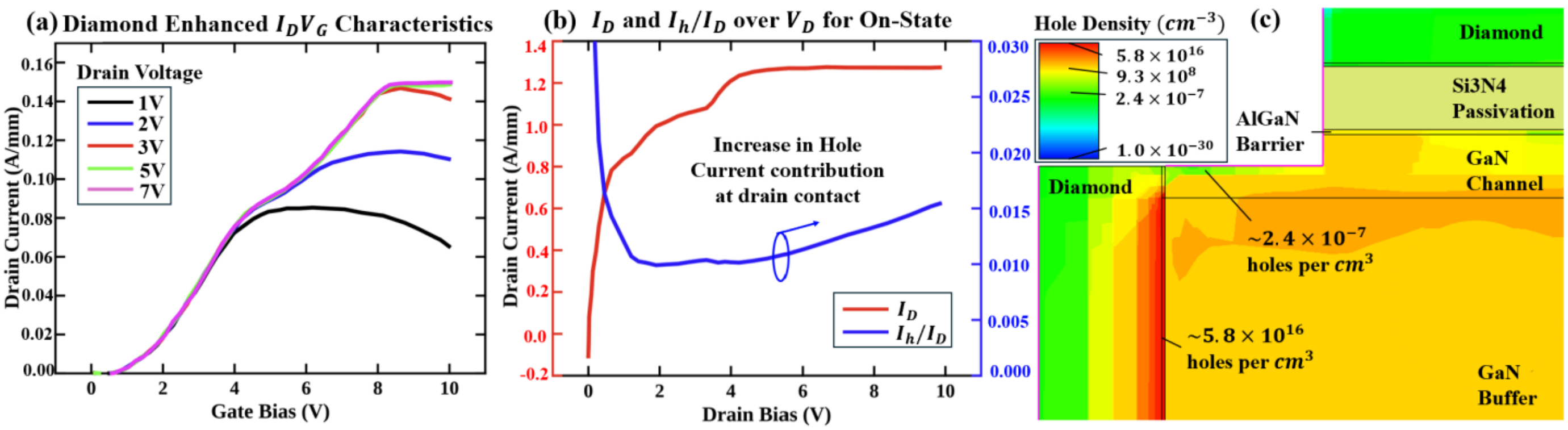

**Fig. 2: (a)** $I_D V_G$ characteristics of 400nm diamond incorporated device. **(b)** $I_D V_D$ contrasted with $I_h/I_D$ shows increase in hole current proportions as On-state is reached, owing to decrease in electron injection. **(c)** Device cross-cut at the source contact definition revealed depletion of holes at the contact and buildup at the diamond interface.

As the interfacial hole accumulation region forms, an additional conduction pathway becomes available near the diamond/GaN interface [9], [10]. Simulations further show reduced hole concentration near portions of the source and drain contacts adjacent to the interface [8], [9]. These effects are consistent with reduced carrier injection efficiency and may contribute to the degradation of drive current observed in the diamond-integrated structure.

## IV. SILVACO DEVICE AND METHODOLOGY

Because the observed behavior originates from interface-driven electrostatics rather than transistor-specific operation [8], [9], the analysis is extended beyond HEMT structures to an isolated Transfer Length Method (TLM) geometry, shown in Fig. 3. This abstraction enables investigation of the underlying interface mechanisms independent of gate control, allowing the influence of diamond integration to be quantified using transport metrics that are not coupled to transistor performance [15]. The simplified geometry provided a controlled platform for isolating leakage pathways and evaluating their dependence on structural implementation [10], [15].

The proposed TLM methodology consists of three primary device configurations. Each structure is formed on uniformly doped n-type GaN and incorporates two electrically identical contacts separated by a lateral spacing $d$. The contacts extend 10 nm into the GaN substrate to suppress artificial corner-field crowding and ensure uniform carrier injection into available conduction paths. The contacts are given a 4.7eV work function to induce Schottky contact behavior. This behavior allows us to measure the mobility of a conductive channel, rather than the bulk resistance. A reference GaN/passivation structure is first used to establish baseline transport behavior and characterize ideal passivation for the given geometry [15].

A direct GaN/diamond configuration is then introduced to evaluate interface-induced conduction associated with high-density hole region formation under varying geometrical conditions [9], [10]. Finally, a passivated GaN/passivation/diamond structure is implemented to investigate mitigation strategies for diamond-enabled leakage, including the influence of dielectric thickness, geometry, and material selection [5], [15]. Together, these structures enable systematic separation of electrostatic interface effects from purely thermal or device-specific behavior [8], [15].

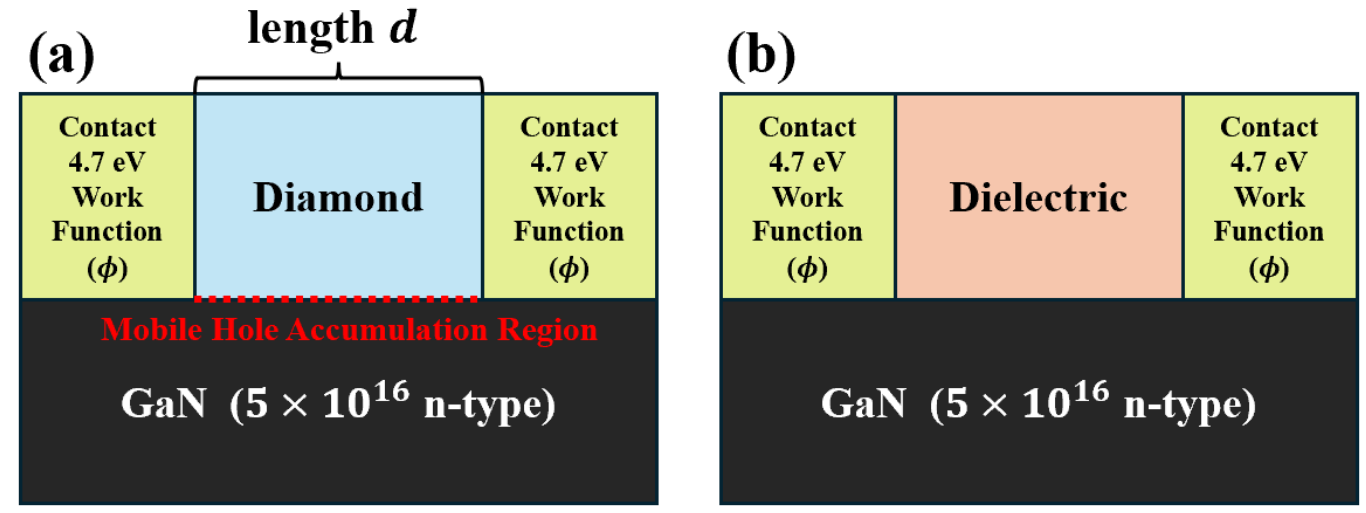

**Fig. 3: (a)** Direct diamond/GaN implementation with anticipated hole accumulation region representation. **(b)** Fully passivated implementation.

## V. INTERLAYER OPTIMIZATION AND CONSIDERATIONS

We first verified the presence of band bending and high-density hole layer formation under zero-bias conditions in the diamond/GaN TLM structure. Analysis of the band profiles revealed electric fields originating at the diamond/GaN interface with magnitudes comparable to those observed in the HEMT simulations. Concurrently, a localized increase in hole concentration was observed near the interface, consistent with band bending sufficient to induce hole accumulation [8], [9].

After replicating the interfacial effect observed in the p-GaN HEMT, we investigated the extent of hole-mediated conduction at the diamond interface using both the fully passivated TLM structure and the direct diamond/GaN TLM. Because the contacts are Schottky in nature, the measured current is dominated by leakage pathways rather than bulk ohmic resistance.

In the fully passivated TLM, the measured saturated current represents the leakage floor associated with conventional dielectric passivation. The resulting current density is extremely low, approximately 2.8×10^(-8) A/mm as shown in Fig. 4(a). As the contact spacing increases, the current decreases monotonically, consistent with suppression of leakage pathways over longer lateral distances.

In contrast, the diamond/GaN TLM exhibits substantially higher leakage current, reaching approximately $5 \times 10^{-2}$ A/mm as shown in Fig. 4(b). Unlike the passivated case, the reduction in current with increasing channel length is non-linear, exhibiting a progressively weaker dependence on spacing. Consequently, diamond-enabled leakage remains significant even for long-channel geometries [15].

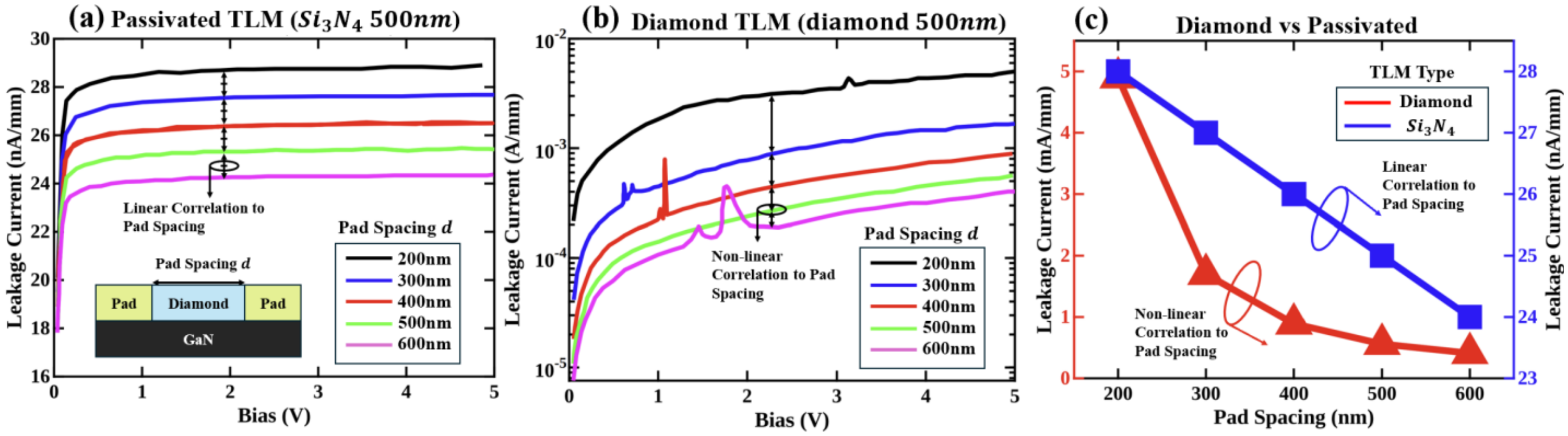

**Fig. 4: (a)** Leakage current as a function of bias for various spacing lengths on the fully passivated device. **(b)** Leakage current as a function of bias for various spacing lengths on the diamond device. **(c)** Approximate leakage as a function of device width, passivated device shows linear decay while the diamond device asymptotically approaches a finite current.

Once the band-bending effects were characterized, we turned our attention to mitigation strategies, using the leakage measurements as a guiding metric. The simplest approach is the introduction of a dielectric interlayer between GaN and diamond [5], [6].

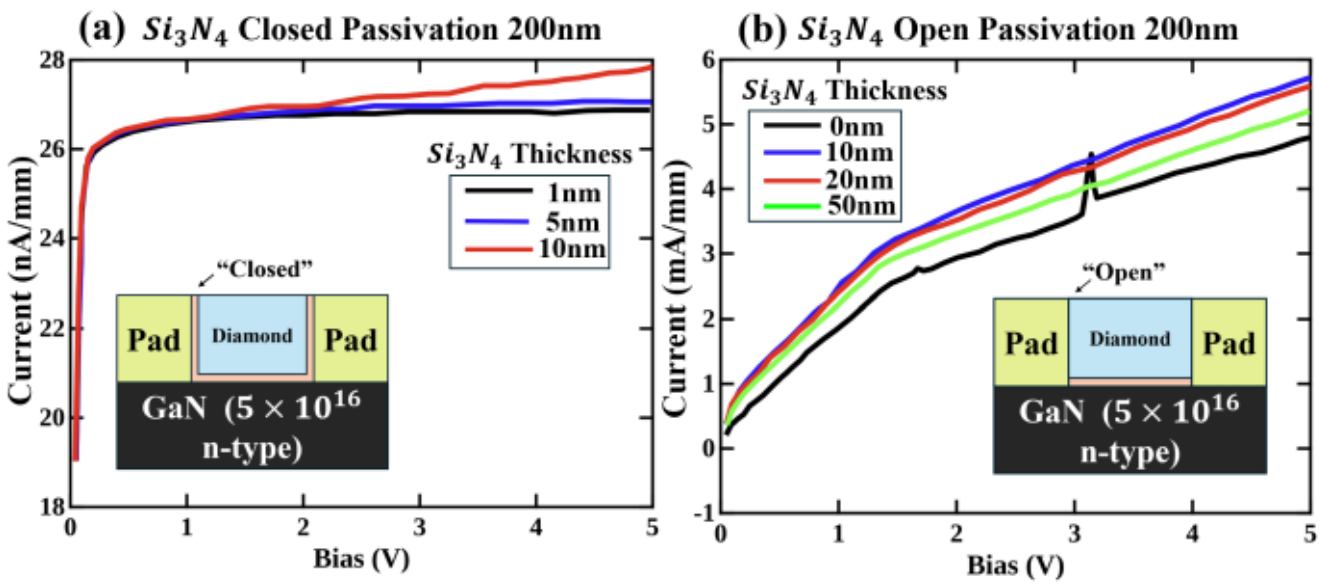

**Fig. 5: (a)** The "Closed" device leakage plots at various GaN to diamond dielectric thickness with 10nm contact sidewall thickness and demonstration of the "Closed" TLM **(b)** The leakage plots for the "open" device at various dielectric thickness and demonstration of "Open" TLM. Note that 0nm is equivalent to removal of the dielectric.

Dielectric-first fabrication processes produce a geometry corresponding to the configuration shown in Fig. 5(c), labeled "Open," and are the first implementation examined. This geometry would arise from a blanket oxide deposition, patterned etch and then deposition of metal. In this structure, the dielectric interlayer is expected to suppress the diamond-induced hole accumulation at the GaN surface, thereby reducing leakage [9], [15]. Contrary to this expectation, the simulated results in Fig. 5(b) showed significant leakage largely independent of passivation thickness or material choice. The dominant current path was found at the diamond/dielectric interface, indicating the formation of an unintended lateral conduction channel within the heat-spreader stack.

The diamond/dielectric/GaN stack introduces a capacitive coupling that enables unintended carrier accumulation at the diamond surface. This effect originates from the large polarization-induced fixed charge at the GaN/dielectric interface [14], [11], which is insufficiently screened when thin dielectric layers are used [15]. The resulting electrostatic imbalance induces an equal and opposite charge at the dielectric/diamond interface, leading to hole accumulation and formation of a conductive channel at the diamond surface [8], [9]. Depending on dielectric thickness and GaN doping concentration, this induced channel can exhibit greater effective current than that observed under direct GaN/diamond contact.

To verify the origin of the observed leakage mechanism, a fourth TLM structure was developed, shown in Fig. 5(c) and labeled "Closed". In this configuration, a 10 nm dielectric layer was introduced along the contact sidewalls to electrically isolate the diamond from the metal contacts. The resulting leakage current was nearly identical to that of the fully passivated reference TLM, indicating that the leakage pathway is enabled primarily through electrical access to the diamond interface rather than the presence of diamond itself. The results demonstrated in Fig. 5(b) strongly support that diamond can be electrically isolated while preserving its thermal advantages, provided that proper sidewall passivation is implemented [5], [6].

This finding represents yet another criticality of diamond integration, indicating that successful implementation requires careful control of contact geometry and dielectric isolation to prevent unintended electrostatic coupling to the diamond surface.

## V. Conclusion

This work validated diamond heat-spreader integration in GaN devices using coupled electrothermal TCAD simulations and dedicated transport test structures [1], [13]. While diamond integration provides substantial thermal benefits and improves high-field device behavior [1], [4], [6], our results demonstrate that it can also introduce electrostatic effects that alter carrier transport [8]. Simulations and Transfer Length Method (TLM) structures revealed that band bending at the diamond/semiconductor interface forms an interfacial hole accumulation layer, creating parasitic conduction pathways independent of the intended channel [8]–[10]. Diamond-induced leakage persisted with increasing device dimensions and exhibited nonlinear channel-width scaling. Passivated GaN/dielectric/diamond structures further demonstrated capacitive coupling between GaN polarization charge and the diamond surface, enabling charge accumulation without direct diamond/GaN contact [8], [11], [15]. Leakage was primarily governed by electrical accessibility of the diamond interface, with dielectric isolation along contact sidewalls effectively suppressing conduction while preserving the thermal pathway [5], [6]. These results highlight an important tradeoff between thermal management and electrostatic integrity in diamond-integrated GaN devices and establish design strategies for maintaining electrical performance while exploiting diamond heat spreading in high-power and RF GaN electronics.